\documentstyle[aas2pp4,psfig,rotate]{article}
\begin{document}


\title{\bf THE OPTICAL-UV CONTINUUM OF A SAMPLE OF QSOs}

\author{F. Natali\altaffilmark{1}, E. Giallongo}
\affil{Osservatorio Astronomico di Roma, via dell'Osservatorio, I-00040,
Monteporzio, Italy}

\author{S. Cristiani} 
\affil{Dipartimento di Astronomia Universit\`a di Padova, vicolo
dell'Osservatorio 5, I-35122 Padova, Italy}

\and

\author{F. La Franca}
\affil{Dipartimento di Fisica - Universit\`a degli studi ``Roma Tre'', 
via della Vasca Navale 84, I-00146 Roma, Italy}

\altaffiltext{1}{Istituto Astronomico - Universit\`a ``La Sapienza''
di Roma,Via Lancisi 29, I-00161 Roma, Italy}

\begin{abstract}
The average optical-UV continuum shape of QSOs has been investigated
using spectra of 62 QSOs having good relative photometric
calibrations. The QSO spectra were extracted from two complete color
selected samples in the magnitude intervals $B \approx 16-20$.  The
analysis was performed fitting power-law continua $f_{\nu}\propto
\nu^{\alpha}$ in well defined rest-frame wavelength intervals after
removing regions of the spectrum affected by strong emission lines or
weak emission bumps.  The average slope in the rest-frame optical-UV
region $1200 - 5500$ {\AA} shows a rapid change around the 3000 {\AA}
emission bump with $\alpha \simeq 0.15$ longward of it and $\alpha
\simeq -0.65$ at shorter wavelengths. Although these average slopes
have been obtained using spectra of QSOs with different luminosities
and redshifts, there are no significant correlations of the average
spectral index with these quantities.  For a few QSOs in the sample we
were able to measure the same softening of the spectral shape within
the individual spectrum. These results have significant consequences
on the estimate of the cosmological evolution of the optically
selected QSOs as they affect, for instance, the $k$-corrections. New
$k$-corrections in the B, V, R and Gr bands were computed.  The
derived average spectral shape in the optical-UV band puts interesting
constraints on the expected emission mechanisms.
\end{abstract}


\section{Introduction}

The study of the shape of the QSO continuum has a great
importance both for the identification of the emission mechanisms
responsible for the observed radiation and for the estimate of the QSO
cosmological evolution.

The continuum spectrum is conventionally parameterized in the
optical-UV region by a power law of the type $f_{\nu}\propto
\nu^{\alpha}$.  This is a local approximation: in the recent years the
availability of IR, UV and X-ray data has allowed to measure the energy
distribution of QSOs in a wide spectral range, showing large
deviations from a single power-law approximation.  Even in the
optical-UV region, composite spectra derived from complete color
selected samples (\cite{crv90,fra91}) show significant
departures from a single power-law.

Another remarkable occurrence is the variety of results obtained in
the many studies of the shape of the optical-UV continuum spectrum, in
particular about the estimated average optical-UV spectral index,
$\alpha$, and its dispersion.  Roughly the intrinsic optical-UV
spectral index ranges from -1 to 0.  O'Brien et al. (1988) and Sargent
et al. (1989) obtain $-0.7 \leq \alpha \leq -0.6$ on two large
samples; Oke \& Korycansky (1982), Oke et al. (1984), Baldwin et
al. (1989) and Cheng et al. (1991), found values in the range $-1 \leq
\alpha \leq -0.5$ from smaller samples. Recently some studies found an
intrinsic spectral index $\alpha\sim -0.3$
(\cite{san89,fra91,web95,fra96}), in agreement with the expected slope
provided by free-free emission models (\cite{bar93}).

However, most of the differences among the results
quoted in the literature might be due to the different spectral
intervals used to estimate the average continuum slope.

No substantial agreement exists also about the dispersion of the
observed spectral index, $\sigma_\alpha$, and its origin: obscuring dust 
(\cite{web95}), intrinsic differences in the spectral energy distribution
from QSO to QSO (\cite{sar89,ser96}), or simply measurement errors 
(\cite{che91}).  Some authors claim no intrinsic dispersion (\cite{san89}), 
others state a large one (\cite{elv94}): so far the observed values of the
dispersion range from 0.2 to 0.5.

Determining the average shape of the optical-UV emission and whether
it is intrinsically the same for all the QSOs is not a worthless
question.  The choice of $\alpha=-0.7$ rather than -0.3 to calculate
the $k$-corrections leads to increase the luminosities of QSOs at
$z\simeq 2$ of a factor $\approx 2$. An appreciable dispersion
($\sigma=0.3$), as well, introduces uncertainties in the luminosity
and biases the estimate of the cosmological evolution of the
luminosity function (\cite{giv92,fra93,laf97}).

In this paper we present a spectroscopic study of a set of 62 objects
extracted from a complete sample of radio-quiet QSOs. We determine the
values of the mean spectral index in four different ranges of rest
frame wavelength, with a particular attention to the correlations
between $\alpha$ and the redshift or the luminosity.

We describe in Sect. 2 the data set and in Sect. 3 the method we used
to perform the power law fit to the continuum. Finally, in Sect. 4, we
show the observed dependence of the spectral index on the mean rest
frame wavelength and present new $k$-corrections to estimate the
evolution of the QSO luminosity function (LF).

\section{Data sample}

The QSOs studied in the present work belong to the Selected Area 94
(\cite{laf92}; $m_B \leq $ 19.9) and the Homogeneous Bright Quasar
Survey (HBQS, \cite{cri95}; $15 \leq m_B \leq $18.75).  Both samples
identified quasar candidates on the basis of UV criteria
(complementary objective-prism data have been also used in a few
subareas to cross check the effectiveness of the selection).  From
this sample a subset of 62 QSOs was extracted with a good relative
photometric calibration and with a luminosity and redshift
distribution similar to the original set (Figure 1).  It is to notice
that the spectral properties of our sample are representative of QSOs
selected with an UV excess and a bias against radio-loud quasars
showing very red colors can be present (\cite{web95,fra96}).  On the
other hand it is not known at present if radio-quiet dust-reddened
objects represent a considerable fraction of the QSO population.

\begin{figure}[ht]
\psfig{figure=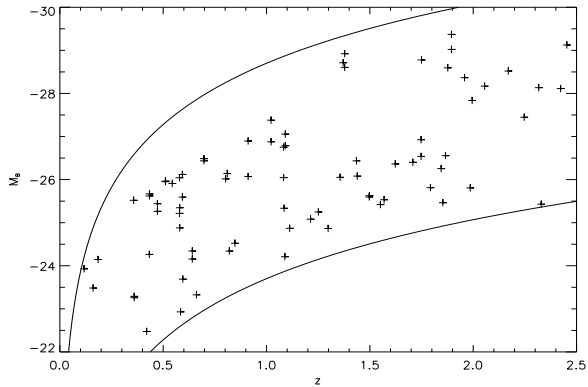,width=8.2truecm,height=5.5truecm}	
\caption{Luminosity {\em vs} redshift plot of the QSO
sub-sample studied. The apparent magnitude loci at $m_B = 15$ and $m_B
= 20$ are shown ($h_0=50$; $q_0=0.5$). \label{fig1}}

\end{figure}

The observations of the QSOs have been carried out at the 3.6m, 2.2m
ESO/MPI and 1.5m ESO telescopes at La Silla, equipped either with the
ESO Faint Object Spectrograph and Camera (EFOSC1 - EFOSC2), or with
the Boller and Chivens Spectrographs.  The slit had typically a width
at least 1.5 times the seeing FWHM and in general was placed at the
parallactic angle. The detectors were always CCDs. The resolution of
the spectra ranged between 10 and 30 \AA.  The reduction process used
the standard MIDAS facilities (\cite{ban83}) available at the Padova
Department of Astronomy and at ESO Garching.

The raw data were sky-subtracted and corrected for pixel-to-pixel
sensitivity variations by division by a suitably normalized exposure
of the spectrum of an incandescent source. Wavelength calibration was
carried out by comparison with exposures of He-Ar, He, Ar and Ne
lamps. Relative flux calibration was finally achieved by observations
of standard stars listed by \cite{oke74} and \cite{sto77}.  The
airmass of the observation was always below 1.6 and has been corrected
on the basis of the standard La Silla extinction tables.

\section{Fitting the continuum}

As discussed in the previous sections the shape of the continuum
component of the QSO spectra changes considerably over different
frequency bands.

In the optical-UV band the continuum is usually parametrized as a
power law: $f_{\nu}\propto \nu^{\alpha}$. Some results
(\cite{crv90,fra91}) show that, even in the optical-UV range
(1200-10000 {\AA}), the continuum spectrum of individual QSOs is not
well represented by a single power law (i.e. $\alpha$ = constant over
the whole band).  However these results are obtained with the
technique of the composite spectra, based on the simplifying
assumption that each spectrum is representative of the whole QSO
``class'' and an ensemble mean can be meaningfully applied.  Such an
hypothesis may result too simplistic and even misleading: it is well
known, for example, that the average of a number of power-laws of
different spectral index is not a single power-law.

In this study we have estimated the {\em local} power-law shape of the
individual QSOs in four ranges, namely the rest-frame intervals: 1400
$-$ 2200 {\AA}, 2150 $-$ 3200 {\AA}, 2950 $-$ 4300 {\AA}, 3900 $-$
5500 {\AA}.  Some spectra of the extracted set of QSOs cover a large
frequency range, so in few cases we have, for the same QSO, two
different rest-frame wavelength intervals. We have obtained
78 spectral ranges from the extracted set of 62 QSOs.

For each spectral range the regions certainly contaminated by strong
emission were eliminated and an automatic algorithm was applied to
estimate the power law index $\alpha$ over the spectral regions
described above. The number of available regions in individual QSO
spectra depends on the redshift and on the instrumental response of
the spectrographs used.

The method consists in a first selection of the probable {\em
continuum window \/} regions within each spectral interval. These
continuum windows are selected on the basis of the composite emission
line spectra published by Cristiani \& Vio (1990) and Francis et
al. (1991).

The local power-law continuum is determined by a linear regression on
the local minima of the spectrum selected after appropriate smoothing
to improve the S/N ratio. Each minimum is weighted by the reciprocal
of the local second derivative of the spectrum, to minimize the
effects of occasional absorptions.  The weight is doubled for those
local minima which, following the results on the composite spectra,
are within identified {\em continuum windows \/} (see Table 1).

\section{Results}

\subsection{Continuum slope}

The mean value of the spectral index, $\overline{\alpha_{\nu}}$, averaged
over the whole sample, is -0.33 with a root mean square dispersion of
0.59, in agreement with the value found by Francis (1993) for the
Large Bright Quasar Survey (LBQS). This value of the dispersion,
however, 

\onecolumn
\begin{figure}[tp]
\epsscale{0.8}
\plotone{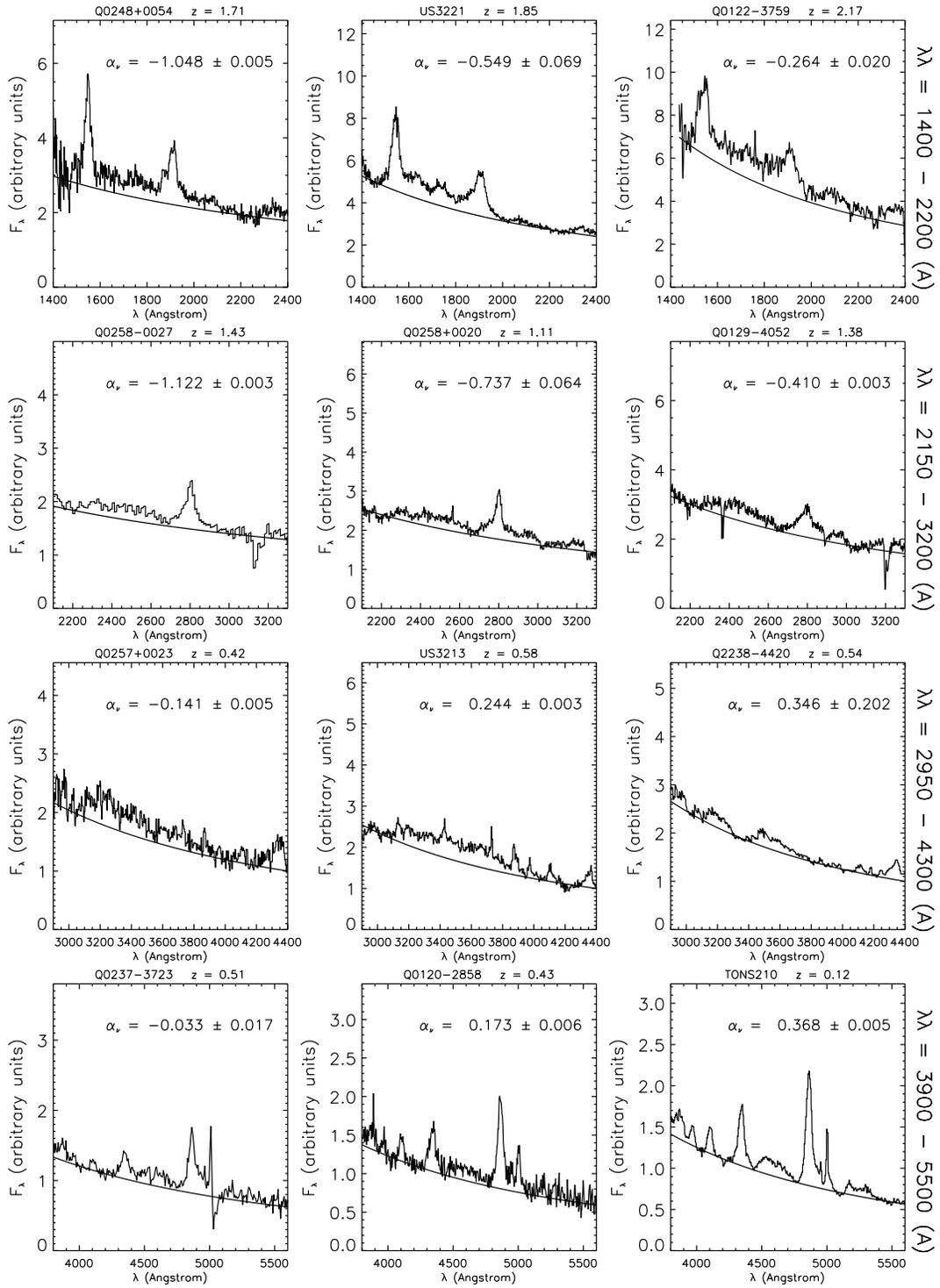}	
\caption{Examples of QSOs' spectra with fitted continuum
and resulting spectral index in the given rest wavelength
interval. \label{fig2}}
\end{figure}
\twocolumn

\noindent
although supported by Elvis et al. (1994), is significantly 
higher than that ($\approx 0.3$) found by
Sanders et al. (1989) and recently by Francis (1996). If the power-law
index is a function of the wavelength, the different results can be
understood in terms of measurements in different rest-frame spectral
regions: indeed, under the hypothesis of significant deviations from
a single power law, we do expect a large dispersion over the whole
sample.

The mean values of the continuum slopes obtained for each of the four
rest wavelength intervals described above are reported in Table 1
together with their dispersions.  For 7 QSOs it was possible to fit
the continuum only in the range 1700 $-$ 2700 {\AA}.
Figure 3 shows the dependence of the spectral index from the mean rest
wavelength of the fitted spectrum: noticeable the presence of a break
in the values from $\overline{\alpha_{\nu}} \simeq -0.65$ to
$\overline{\alpha_{\nu}} \simeq 0.15$ at $\overline{\lambda} \simeq 3200$
{\AA}.  In this case the dispersions are lower (see Table 1) than
found for the whole sample.

\begin{figure}[ht]
\psfig{figure=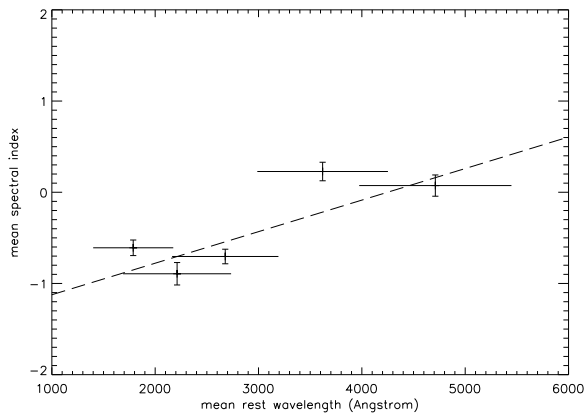,width=8.2truecm}	
\caption{Dependence of the mean spectral index from the
mean rest wavelength of the fitted spectrum. Along the y-axis are
plotted the errors in $\overline{\alpha}$, while along the x-axis the
average baseline of the fit is shown for each rest wavelength
interval. \label{fig3}}
\end{figure}

To represent the overall shape of the average QSO continuum as a
gradual change of the local slope as a function of wavelength we have
assumed a linear trend of $\alpha_{\nu}$ vs $\overline{\lambda}_{rest}$
(Figure 3),
i.e. $\overline{\alpha_{\nu}} = a ($${\overline{\lambda}} \over 4400$$) + b$
(where the wavelength is expressed in Angstrom); we derived $a = 1.5
\pm 0.3$ and $b = -1.5 \pm 0.2$ by a linear fit with errors in both
coordinates (see \cite{fav88}). In this case, because of  
$\alpha_{\nu} = \frac{dLog F_{\nu}}{dLog {\nu}}$, the
optical-UV continuum is expressed by a more complicated function of
the wavelength:

\begin{equation}
Log F_{\nu} = \left( b - \frac{k}{\nu Log\nu} \right) Log\nu
\end{equation}

\bigskip

where $ k = \frac{a}{4400 ~ ln10} \; c $ ($c$ is the speed of light).

No significant correlation between $\alpha_{\nu}$ and the redshift or the
luminosity was found (Table 2), in agreement with the results obtained
by Richstone \& Schmidt (1980), Oke et al. (1984) and Cheng et
al. (1991).  In all the four wavelength ranges we have $\mid
\frac{d\alpha_{\nu}}{dz} \mid < 0.4 \pm 0.4$ and 
$\mid \frac{d\alpha_{\nu}}{dM_B}
\mid < 0.1 \pm 0.1$, with a probability greater than 15\% of no
correlation between $\alpha_{\nu}$ and redshift or luminosity. Since
$\overline{\alpha_{\nu}}$ has been found roughly constant below and above
3200 {\AA}, we studied the correlation with $z$ and $M_{B}$ in the two
larger wavelength intervals $1400 - 3200$ {\AA} and $3200 - 5500$
{\AA} (Figure 4), combining the values of $\alpha$ found in each of the
intervals reported in Table 1: even in this case no significant 
correlation was found.

To avoid spurious results due to the luminosity-redshift correlation
present, to some degree, in flux-limited QSO samples, we repeated the
correlation analysis in restricted intervals of luminosity and
redshift.  The mean values of $\alpha_{\nu}$ and their degree of correlation
with $z$ and $M_{B}$ do not change substantially if we choose objects
of the sample having luminosities or redshifts in the following
intervals: $-27.5 \leq M_{B} \leq -25.5$, $ 0.5 \leq z \leq 1.1$.
These results suggest that the variations of the spectral index with
$\overline{\lambda}_{rest}$ are intrinsic and do not depend on the
distribution of QSOs in the redshift-luminosity plane.

As a further check for thirteen objects it was possible to measure the
change in the average continuum slope around 3000 {\AA} by fitting the
individual continua over two different (contiguous) wavelength baselines: five
QSOs over $\lambda\lambda 2150 - 3200$ {\AA} and $\lambda\lambda
2950 - 4300$ {\AA}, eight over $\lambda\lambda 2950 - 4300$
{\AA} and $\lambda\lambda 3900 - 5550$ {\AA}.  The resulting mean
difference in the
spectral indices between the two spectral intervals, averaged over each 
of the two sets of QSOs, are, in both cases, close to the difference in the 
values of the mean spectral indices found for the whole sample.

\onecolumn
\begin{table}[hp]
\begin{center}
\begin{tabular}{ccccc}
\tableline
rest wavelength interval & continuum windows & 
$\overline{\alpha}_{\nu}$ &$\sigma$ & number of QSOs\\
\tableline\tableline
1400-2200 {\AA} & \begin{tabular}{c}1420-1480 {\AA}\\
				    2150-2300 {\AA} \end{tabular} &
-0.61 & 0.37 & 19\\
\tableline
{\em 1700-2700 {\AA}} & \begin{tabular}{c}{\em 2150-2300 {\AA}}\\
					  {\em 2600-2700 {\AA}} \end{tabular} &
{\em -0.89} & {\em 0.32} & {\em 7}\\
\tableline
2150-3200 {\AA} & \begin{tabular}{c}2150-2300 {\AA}\\
				    2600-2700 {\AA}\\
				    2950-3100 {\AA} \end{tabular} &
-0.70 & 0.35 & 19\\
\tableline
2950-4300 {\AA} & \begin{tabular}{c}2950-3100 {\AA}\\
				    3900-4300 {\AA} \end{tabular} &
0.23 & 0.48 & 22\\
\tableline
3900-5500 {\AA} & \begin{tabular}{c}3900-4300 {\AA}\\
				    5400-5500 {\AA} \end{tabular} &
0.07 & 0.39 & 11\\
\tableline
\end{tabular}
\end{center}

\caption{Mean values and dispersions of the continuum slopes obtained for 
each of the four rest wavelength intervals in which the variations of the 
spectral index $\alpha$ are within the errors of the fit; also the result for 
the interval 1700 $-$ 2700 {\AA} (7 QSOs) is shown. In the second column are
reported the {\em continuum window \/} 
regions (identified following the results on the composite spectra) falling in
each rest wavelength interval.\hfill \label{tab1}}
\end{table}

\vspace{2.5truecm}

\begin{table}[hp]
\begin{center}
\begin{tabular}{ccccccc}
\tableline
rest wavelength interval & 1400-2200 {\AA} & 1700-2700 {\AA} &
2150-3200 {\AA} & 2950-4300 {\AA} & 3900-5500{\AA}\\
\tableline\tableline
$\frac{d\alpha_{\nu}}{dz}$ & $0.43 \pm 0.36$ & ${-}$ & $-0.11 \pm 0.27$ &
$0.42 \pm 0.47$ & $0.30 \pm 0.63$\\
$r_{\alpha,z}$ & 0.277 & ${-}$ & 0.101 & 0.199 & 0.158\\ 
$P(r_{\alpha,z})$ & $0.25$ & $-$ & $0.68$ & $0.37$ & $0.64$\\
\tableline
$\frac{d\alpha_{\nu}}{dM_B}$ & $0.07 \pm 0.07$ & ${-}$ & $-0.02 \pm 0.06$ &
$-0.07 \pm 0.08$ & $-0.06 \pm 0.11$\\
$r_{\alpha,M_B}$ & 0.246 & ${-}$ & 0.069 & 0.181 & 0.171\\
$P(r_{\alpha,M_B})$ & 0.31 & ${-}$ & 0.78 & 0.42 & 0.61\\  
\tableline\tableline 
$\frac{d\alpha_{\nu}}{dz}$ & \multicolumn{3}{c}{$0.11 \pm 0.12$} &
\multicolumn{2}{c}{$0.45 \pm 0.33$}\\
$r_{\alpha,z}$ & \multicolumn{3}{c}{0.146} & \multicolumn{2}{c}{0.239}\\
$P(r_{\alpha,z})$ & \multicolumn{3}{c}{0.34} & \multicolumn{2}{c}{0.18}\\
\tableline
$\frac{d\alpha_{\nu}}{dM_B}$ & \multicolumn{3}{c}{$-0.00 \pm 0.04$} &
\multicolumn{2}{c}{$-0.07 \pm 0.06$}\\
$r_{\alpha,M_B}$ & \multicolumn{3}{c}{0.013} & \multicolumn{2}{c}{0.187}\\ 
$P(r_{\alpha,M_B})$ & \multicolumn{3}{c}{0.93} & \multicolumn{2}{c}{0.30}\\
\tableline

\end{tabular}
\end{center}

\caption{Correlation between the spectral index $\alpha_{\nu}$ and both
the redshift and the luminosity; $r$ is the correlation coefficient and
$P(r)$ is the probability of no correlation. The absolute magnitude was
calculated assuming $h_0 = 50$ and $q_0 = 0.5$. \label{tab2}}
\end{table}

\twocolumn

\begin{figure}[ht]
\twocolumn[{
\begin{tabular}{c}
\rotate[l]{\psfig{figure=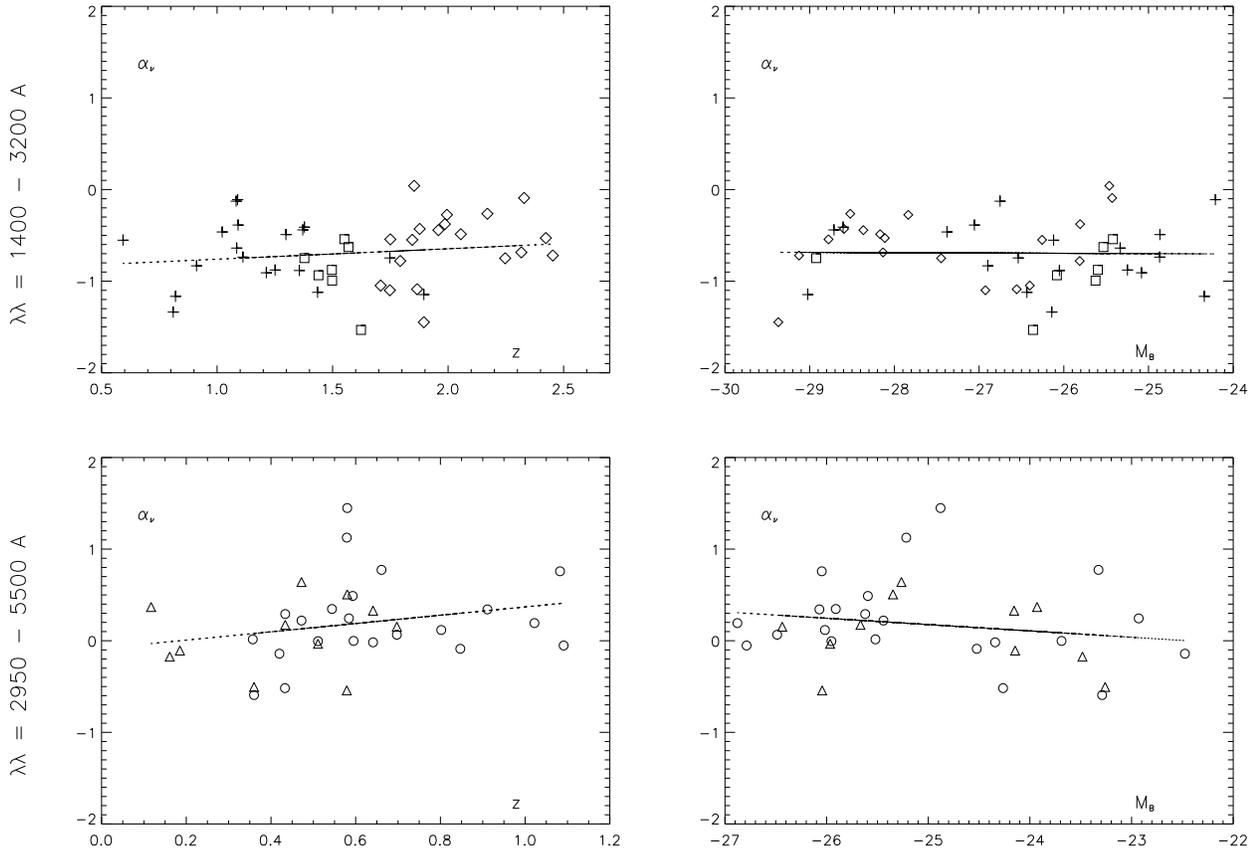,width=12truecm}}\\
\end{tabular}
		
\caption{Correlation between the spectral index and both
the redshift ({\em left}) and the luminosity ({\em right}). Different
symbols indicate values of $\alpha$ belonging to different rest-frame
intervals: 1400 $-$ 2200 ({\em diamond}); 1700 $-$ 2700 ({\em
square}); 2150 $-$ 3200 ({\em cross}); 2950 $-$ 4300 ({\em circles});
3900 $-$ 5500 ({\em triangles}). Here is assumed $h_0=50$ and
$q_0=0.5$. \label{fig4}}
}]
\end{figure}

\subsection{$k$-corrections and the evolution of the
QSO luminosity function}

The shape of the continuum spectrum has an important role in
determining the $k$-corrections and therefore the luminosity function
of the QSOs. In fact, because of the different redshifts of the
observed objects, to evaluate properly the QSO monochromatic
luminosity function and its evolution, the observed flux of the QSOs
has to be corrected by a quantity, the $k$-correction, that depends on
the shape of the spectrum.

The $k$-correction is given by the usual formula

\begin{equation}
k = 2.5 log(1+z) + 2.5 log
\frac{ {\int_{0}^{\infty}} F(\lambda)s(\lambda)d\lambda }
{ {\int_{0}^{\infty}} F(\frac{\lambda}{1+z}) s(\lambda)d\lambda }
\end{equation}

\bigskip

$K$-corrections in the Johnson's B, V, R and Gunn's Gr bands (Table 3) 
were calculated using equation (2) with
a {\em composite spectrum} (Figure 7) built as follows: we computed
the continuum spectrum from equation (1) and the emission component
averaging the residuals, respect to their own continuum, of each
spectrum of the sample.  The $k$-corrections in the blue bandpass were
compared (Figure 5) with the ones derived from two composite spectra
having a power law continuum with constant spectral indexes,
respectively $\alpha = -0.3$ and $\alpha = -0.5$ and with the $k$-corrections
given by Cristiani \& Vio (1990). 

\onecolumn

\begin{table}[ht]
\begin{center}
\begin{tabular}{rrrrrrrrrrrrrrr}
\tableline 
 & z & & & $k_{B}$ & & & $k_{V}$ & & & $k_{R}$ & & & $k_{Gr}$ &  \\ 
\tableline \tableline
 & 0.1 & & & -0.087 & & & -0.144 & & & -0.036 & & &  0.133 & \\
 & 0.2 & & & -0.179 & & & -0.234 & & & -0.125 & & &  0.242 & \\
 & 0.3 & & & -0.232 & & & -0.289 & & & -0.215 & & &  0.051 & \\
 & 0.4 & & & -0.298 & & & -0.367 & & & -0.281 & & & -0.084 & \\
 & 0.5 & & & -0.348 & & & -0.443 & & & -0.342 & & & -0.118 & \\
 & 0.6 & & & -0.391 & & & -0.490 & & & -0.408 & & & -0.146 & \\
 & 0.7 & & & -0.420 & & & -0.522 & & & -0.471 & & & -0.202 & \\
 & 0.8 & & & -0.425 & & & -0.589 & & & -0.520 & & & -0.286 & \\
 & 0.9 & & & -0.434 & & & -0.623 & & & -0.549 & & & -0.350 & \\
 & 1.0 & & & -0.455 & & & -0.661 & & & -0.600 & & & -0.384 & \\
 & 1.1 & & & -0.506 & & & -0.689 & & & -0.646 & & & -0.390 & \\
 & 1.2 & & & -0.529 & & & -0.697 & & & -0.672 & & & -0.409 & \\
 & 1.3 & & & -0.542 & & & -0.698 & & & -0.703 & & & -0.478 & \\
 & 1.4 & & & -0.554 & & & -0.711 & & & -0.727 & & & -0.532 & \\
 & 1.5 & & & -0.581 & & & -0.726 & & & -0.740 & & & -0.545 & \\
 & 1.6 & & & -0.610 & & & -0.777 & & & -0.743 & & & -0.549 & \\
 & 1.7 & & & -0.604 & & & -0.811 & & & -0.752 & & & -0.561 & \\
 & 1.8 & & & -0.600 & & & -0.829 & & & -0.761 & & & -0.569 & \\
 & 1.9 & & & -0.591 & & & -0.839 & & & -0.780 & & & -0.559 & \\
 & 2.0 & & & -0.581 & & & -0.848 & & & -0.819 & & & -0.552 & \\
 & 2.1 & & & -0.585 & & & -0.862 & & & -0.838 & & & -0.563 & \\
 & 2.2 & & & -0.630 & & & -0.903 & & & -0.848 & & & -0.588 & \\
 & 2.3 & & & -0.659 & & & -0.917 & & & -0.855 & & & -0.628 & \\
 & 2.4 & & & -0.621 & & & -0.913 & & & -0.863 & & & -0.686 & \\
 & 2.5 & & & -0.515 & & & -0.915 & & & -0.872 & & & -0.716 & \\
 & 2.6 & & & $-$ & & & $-$ & & & -0.897 & & & -0.717 & \\
 & 2.7 & & & $-$ & & & $-$ & & & -0.928 & & & -0.704 & \\
 & 2.8 & & & $-$ & & & $-$ & & & -0.930 & & & -0.682 & \\
 & 2.9 & & & $-$ & & & $-$ & & & -0.923 & & & -0.675 & \\
 & 3.0 & & & $-$ & & & $-$ & & & -0.921 & & & -0.689 & \\
 & 3.1 & & & $-$ & & & $-$ & & & -0.919 & & & -0.726 & \\
 & 3.2 & & & $-$ & & & $-$ & & & -0.914 & & & -0.774 & \\
 & 3.3 & & & $-$ & & & $-$ & & & -0.908 & & & -0.788 & \\
 & 3.4 & & & $-$ & & & $-$ & & & -0.901 & & & -0.772 & \\
 & 3.5 & & & $-$ & & & $-$ & & & -0.910 & & & -0.748 & \\
 & 3.6 & & & $-$ & & & $-$ & & & -0.952 & & & -0.721 & \\
 & 3.7 & & & $-$ & & & $-$ & & & -0.986 & & & -0.691 & \\
 & 3.8 & & & $-$ & & & $-$ & & & -0.990 & & & -0.671 & \\
 & 3.9 & & & $-$ & & & $-$ & & & -0.960 & & & -0.670 & \\
 & 4.0 & & & $-$ & & & $-$ & & & -0.891 & & & -0.683 & \\

\end{tabular}
\end{center}
\caption{$K$-corrections in the B, V, R (Johnson) and Gr (Gunn) 
bandpasses as derived from our composite spectrum. \label{tab3}}

\end{table}

\twocolumn

It is to notice that for $z<1$ there is a good agreement between our
$k$-corrections and the ones derived from a spectrum with a constant
slope $\alpha_{\nu} = -0.3$, whereas the difference rises to about 0.2
magnitudes at $z \sim 2$. This suggests that the hypothesis of a flat
single-power-law spectrum leads to an underestimate of the luminosity
of the QSO at high redshifts. Our results are in better agreement
with the $k$-correction computed by Cristiani \& Vio (1990) although
in this case lower luminosities are predicted at $z < 2$ by our 
curve.

\begin{figure}[ht]
\psfig{figure=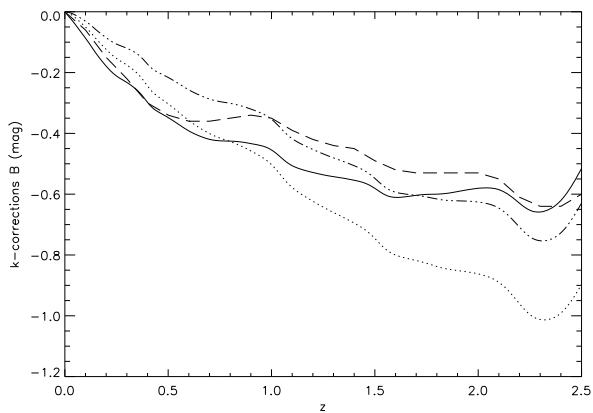,width=8.2truecm}	
\caption{$K$-corrections in the B bandpass calculated in
the case of our synthetic spectrum ({\em solid}) and spectra having
single power-law continuum with spectral indexes -0.3 ({\em dotted})
and -0.5 ({\em dot-dashed}) compared with $k$-corrections given by
Cristiani \& Vio (1990) ({\em dashed}). \label{fig5}}
\end{figure}

This new $k$-correction was used to estimate, by a maximum likelihood
method, the optical luminosity function $\Phi(M_{B},z)$ for a large
sample of $\sim 1000$ QSOs, taken from various surveys in the literature
as described in La Franca \& Cristiani (1997). A double power law 
luminosity function 
and a pure
luminosty evolution (PLE) model of the type (see \cite{laf97} for details):

\begin{flushleft}
$\Phi(M_{B},z) =$\\
\end{flushleft}
\begin{eqnarray}
= \frac{\Phi^{*}}
{10^{0.4\left[M_{B}-M_{B}^{*}(z)\right]
\left(\alpha+1\right)}
+ 10^{0.4\left[M_{B}-M_{B}^{*}(z)\right]\left(\beta+1\right)}}
\end{eqnarray}\\

\bigskip
were assumed, where $\alpha$ and $\beta$ are the slopes of the
faint-end and bright-end luminosity function, respectively.
$\Phi(M_{B},z)$ is in units of Mpc$^{-3}$mag$^{-1}$. In this model the
evolution is specified by the redshift dependence of the break
magnitude ($M_{B}^{*}(z) = M_{B}^{*} - 2.5 k log(1+z)$), where
$M_{B}^{*}\equiv M_{B}^{*}($z=0$)$. This relationship corresponds to a
power law evolution of the luminosity:

\begin{equation}
L^{*}(z) = L^{*} (1+z)^{\kappa}
\end{equation}

\bigskip

The results are shown in Table 4 and Figure 6. The luminosity function
computed using our $k$-correction has a value of the faint-end slope
$\alpha$ between the values of the faint-end slope derived from the
single power-law spectra, while the bright-end slope $\beta$ is the
same (within the errors). Moreover, the $z=0$ break luminosity
($M_B^{*} \simeq -22.5$) does not change substantially with respect to
the LF computed with the assumption of a single power-law slope, but
the parameter $\kappa$, which gives the rate of the cosmological
evolution, goes from the value 2.90 for a spectral index -0.3 to 3.12
for our spectrum, and the probability of the fit improves noticeably
(Table 4).

Comparing our luminosity function with that computed by La Franca \&
Cristiani (1997) using the Cristiani \& Vio (1990) $k$-correction we 
confirm the flattening at low redshift of the
bright part of the luminosity funtion in comparison with a Pure Luminosity
Evolution model. This is indicated by the poor fitting probabilities
given by the two dimensional Kolmogorov-Smirnov test in the range
($0.3\leq z \leq 0.6$) (see Table 4). However, the larger differences
between our and Cristiani \& Vio (1990) $k$-correction are at $z\sim2$, where
the discrepancy is of 0.2 magnitudes. This is at the origin of a
evolutionary parameter which is about 2$\sigma$ lower,
resulting in a volume density of QSOs at $z\sim 2$ which is lower
by a factor about 1.5 at $M_B\sim -27.5$.

Our more accurate estimates of the dispersions of the average spectral
index as a function of wavelength ($\sigma =0.3-0.4$ from Table 1) can
also be included in the evaluation of the evolution of the luminosity
function.  Uncertainties in the estimate of the average spectral
slopes are known to produce a smaller cosmological evolution of the
luminosity function (\cite{giv92}). In this case a value
for the evolutionary parameter as small as $k=2.94$ has been found (see
Table 4). The probability of the PLE evolutionary models with the new
$k$-corrections are in general higher than previously found ($P(\chi
^2)\sim 0.5$), showing how the selection of a more accurate average
spectral shape can discriminate between different evolutionary models
for the QSO luminosity function. In particular, there is no evidence
for any redshift cutoff in the redshift evolution of the luminosity
function as introduced by Boyle (1992) at $z=1.9$.

\begin{figure}[h]
\plotone{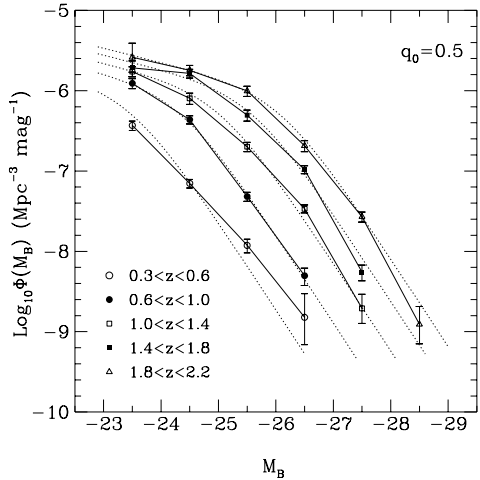}	
\caption{The QSO luminosity function. The points
connected with a continuous line represent the observations (see La
Franca \& Cristiani, 1997 for details); the dotted lines are the
best-fitting Pure Luminosity Evolution model.  Error bars correspond
to $1\sigma$ confidence intervals. \label{fig6}}
\end{figure}

\begin{table}[ht]
\twocolumn[{
\begin{center}
\begin{tabular}{lcccccccc}
\tableline 
\multicolumn{2}{c}{spectrum model} & $\kappa$ & $\beta$ & $\alpha$ & $M^{*}$
& $\Phi^{*}$ & $P(\chi^{2})$ & $P$(K-S) \\
 & & & & & $mag^{-1}Mpc^{-3}$ & $(z < 2.2)$ & $(0.3< z < 0.6)$ \\
\tableline\tableline
$\alpha_{\nu} = -0.3$ & (a) &
2.90 & -3.79 & -1.51 & -22.5 & $8.9 \times 10^{-7}$ & 0.18 & 0.007  \\
$\alpha_{\nu} = -0.5$ & (a) &
2.97 & -3.84 & -1.74 & -22.9 & $5.6 \times 10^{-7}$ & 0.30 & 0.017 \\
Cristiani \& Vio (1990) & (a) &
3.26 & -3.72 & -1.39 & -22.3 & $1.1 \times 10^{-6}$ & 0.21 & 0.02 \\
$\alpha_{\nu} = 1.5(\frac{\lambda}{4400}) - 1.5$ & (a) &
3.12 & -3.73 & -1.44 & -22.4 & $9.6 \times 10^{-7}$ & 0.56 & 0.013 \\
$\alpha_{\nu} = 1.5(\frac{\lambda}{4400}) - 1.5$ & (b) &
2.94 & -3.84 & -1.47 & -22.4 & $1.0 \times 10^{-6}$ & 0.30 & 0.005 \\
\tableline
\multicolumn{2}{c}{$1\sigma$ errors} & $\pm 0.07$ & $\pm 0.13$ & $\pm 0.07$ & $\pm 0.2$ & & &  \\

\end{tabular}
\end{center}

\caption{Parameters for luminosity function with different
continuum shapes: (a) PLE model ($q_0=0.5$); (b) PLE model 
including a dispersion in the distribution of the spectral slope
$\sigma = 0.4$ ($q_0=0.5$).
The last two columns are respectively the $\chi^2$ test
probability and the two-dimensional Kolmogorov-Smirnov test probability
(see details in La Franca \& Cristiani, 1997). \label{tab4}}
\vspace{1truecm}}]
\end{table}

\subsection{Modelling the emission mechanism} 

As an example of the consequences that more extended works of
this type can have in constraining thermal emission models for the
QSO continuum, we have compared with thermal models based on the
presence of an accretion disk around a massive Kerr black hole
($10^6-10^9$ M$_{\odot}$) (\cite{cze87}). These models have
been already used to fit the continua of QSOs in a wide wavelength
interval extending to the X-ray band (\cite{fio95}).

The emission comes from a geometrically thin accretion disk around a
massive black hole with accretion rates $\dot{m}$ ranging in the
sub-Eddington regime. For accretion rates much higher than the
Eddington one the disks become slim and additional cooling processes
should be included. In this illustrative example only sub-Eddington
thin disks are considered.

\begin{figure}[ht]
\psfig{figure=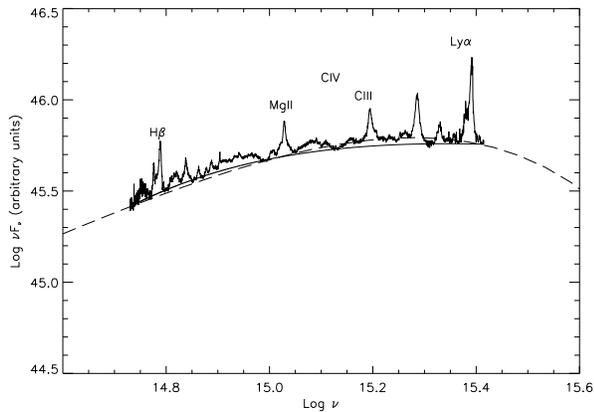,width=8.2truecm}	
\caption{Composite spectrum with the continuum component
({\em solid}), as estimated in the present work, compared to the
synthetic continuum spectrum derived from thermal emission models of
an accretion disk around a massive Kerr's black hole.  The best
fitting model ($M = 10^{9.5} M_{\odot}$, $\dot{m} = 0.8$ and
$cos\theta = 0.75$; see Fiore et al., 1995, for details) is shown
({\em dashed}). \label{fig7}}
\end{figure}

The average shape has been normalized to $M_B\sim -24$ since higher
luminosities require emission in a super-Eddington regime which is not
allowed by the model. A qualitative comparison seems to indicate that
only models with a large black hole mass ($M \simeq 10^{9.5}
M_{\odot}$), an average inclination along the line of sight $\mu=cos
\theta=0.5$ with an intermediate accretion rates $\dot{m}=0.3$, or smaller
inclination ($\mu=cos\theta=0.75$) and higher accretion rate
($\dot{m}=0.8$) can reproduce a change of slope $\Delta\alpha \approx 1$ at
$\overline{\lambda} \approx 3200$ {\AA} (Figure 7). Models with
smaller black hole masses ($M \leq 10^{7.5} M_{\odot}$) provide much
lower luminosities, while models with masses between $10^{8}
M_{\odot}$ and $10^{8.5} M_{\odot}$ present a softer continuum shape
than the observed one.  For black hole masses $M \approx 10^{9} -
10^{9.5} M_{\odot}$, high accretion rates, and smallest inclination
angles ($cos \theta=1$ corresponds to a face-on disk) there is a good
agreement for $Log\nu \leq 15.2$, but the large masses involved in the
accretion process produce a drop of the luminosity in the blue part of
the spectrum.

\section{Conclusion}

In this paper we have analysed the spectra of 62 QSOs in the
magnitude interval $B=15-20$ having good relative photometric
calibrations.  The analysis was performed fitting power-law continua
$f_{\nu}\propto \nu^{\alpha}$ in well defined rest-frame wavelength
intervals after removing regions of the spectrum affected by strong
emission lines or weak emission bumps.  

The main results of the paper can be summarized as follows:

\begin{itemize}
\item
The mean values of the continuum slopes obtained in the spectra of 62 QSOs
change from $\alpha \sim 0.15$ at $\lambda >3000$ {\AA} to $\alpha \sim -0.65$
at $\lambda < 3000$ {\AA}.
\item
Assuming a linear change of the local slope with the rest-frame wavelength
we obtain the following relation $\alpha = 1.5 (\lambda/4400) -1.5$
\item
There is no correlation in the sample between $\alpha$ and redshift and/or
luminosity.
\item
New $k$-corrections, which include the contribution of emission
features, are derived for the B, V, R (Johnson) and Gr (Gunn) bands.
\item
The new $k$-corrections are used to estimate the evolution of the QSO
luminosity function. Assuming PLE models, the double power-law
luminosity function evolves as $L\propto (1+z)^{3.1}$ or as low as
$L\propto (1+z)^{2.9}$ if we take into account the dispersion in the
distribution of spectral shapes.
\item
A qualitative comparison of the QSO average continuum shape with
thermal models based on the presence of a thin accretion disk around a
massive Kerr black hole seems to indicate that only models with a
large black hole mass ($M \simeq 10^{9.5} M_{\odot}$) can reproduce a
change of slope $\Delta\alpha \approx 1$ at $\overline{\lambda}
\approx 3200$ {\AA}.

\end{itemize}

\acknowledgments

The authors acknowledge B. Czerny and
A. Siemiginowska for the code of the accretion disk model and F. Fiore
for useful discussions. We warmly thank the referee P. Francis for his
suggestions. This research has made use of the Simbad database,
operated at CDS, Strasbourg, France.


\begin{thebibliography}{}
\bibitem[Baldwin et al.]{bal89} Baldwin, J. A., Wampler, W. J. \& 
	Gaskell, C. M. 1989 \apj, 338, 630 
\bibitem[Banse et al. 1983]{ban83} Banse, K., Crane, P., Ounnas, C. \& 
	Ponz, D. 1983 In: {\it Proc. of DECUS}, Zurich, p. 87 
\bibitem[Barvainis 1993]{bar93} Barvainis, R. 1993, \apj, 412, 513
\bibitem[Boyle 1992]{boy92} Boyle, B. J. 1992 in ``Texas/ESO-CERN Symposium 
	on Relativistic
	Astrophysics, Cosmology and Particle Physics'', ed(s) Barrow et al.,
	Ann. N.Y. Acad. of Sci., 647, 14
\bibitem[Boyle et al. 1988]{boy88} Boyle, B. J., Shanks, T. \& 
	Peterson, B. A. 1988, MNRAS, 235, 935 
\bibitem[Cheng et al. 1991]{che91} Cheng, F. H., Gaskell, C. M. \& 
	Koratkar, A. P. 1991, \apj, 370, 487 
\bibitem[Cristiani et al. 1995]{cri95} Cristiani, S., La Franca, F., Andreani, 
	P., Gemmo, A., Goldschmidt, P., 
	Miller, L., Vio, R., Barbieri, C., Bodini, L., Iovino, A., Lazzarin, 
	M., Clowes, R., MacGillivray, H., Gouiffes, Ch., Lissandrini, C.  
	\& Savage, A. 1995, A\&AS, 112, 347
\bibitem[Cristiani et al. 1996]{cri96} Cristiani, S., Trentini, S., La Franca, 
	F., Aretxaga, I., Andreani,
	P., Vio, R., \& Gemmo, A. 1996, A\&A, 306, 395 
\bibitem[Cristiani \& Vio 1990]{crv90} Cristiani, S., \& Vio, R. 1990, 
	A\&A, 227, 385 
\bibitem[Czerny \& Elvis 1987]{cze87} Czerny, B., \& Elvis, M. 1987,
	\apj, 321, 305 
\bibitem[Elvis et al. 1994]{elv94} Elvis, M., Wilkes, B. J., McDowell, J. C., 
	Green, 
	R. F., Bechtold, J., Willner, S. P., Oey, M. S., Polomsky, E., \& 
	Cutri, R. 1994, \apjs, 95, 1 
\bibitem[Fasano \& Vio 1988]{fav88} Fasano, G. \& Vio, R. 1988, 
	``Newsletter of the Working Group for Modern
	Astronomical Methodology'' 7, 2 
\bibitem[Fiore et al. 1995]{fio95} Fiore, F., Elvis, M., Siemiginowska, A., Wilkes, B. J.,
	McDowell, J. C.,  \& Mathur, S. 1995, \apj, 449, 74 
\bibitem[Francis et al. 1991]{fra91} Francis, P. J., Hewett, P. C., 
	Foltz, C. B., Chaffee, F. H., Weymann,
	R. J., \& Morris, S. L. 1991, \apj, 373, 465 
\bibitem[Francis 1993]{fra93} Francis, P. J. 1993, \apj, 407, 519
\bibitem[Francis 1996]{fra96} Francis, P. J. 1996, PASA, 13, 212 
\bibitem[Giallongo \& Vagnetti 1992]{giv92} Giallongo, E., \&
	Vagnetti, F. 1992, \apj, 396, 411
\bibitem[La Franca et al. 1992]{laf92} La Franca, F., Cristiani, S., \& 
	Barbieri, C. 1992, AJ, 103, 1062 
\bibitem[La Franca \& Cristiani 1997]{laf97} La Franca, F., \& Cristiani, S. 1997, 
	AJ, 113, 1517
\bibitem[O'Brien et al. 1988]{obr88} O'Brien, P. T., Gondhalekar, P. M., \& 
	Wilson, R. 1988, MNRAS, 233, 801 
\bibitem[Oke 1974]{oke74} Oke, J. B. 1974, \apjs, 27, 21 
\bibitem[Oke \& Korycansky 1982]{oke82} Oke, J. B., \& Korycansky, D. G. 
	1982, \apj, 255, 11 
\bibitem[Oke et al. 1984]{oke84} Oke, J. B., Shields, G. A., \& 
	Korycansky, D. G. 1984, \apj, 277, 64 
\bibitem[Richstone \& Schmidt 1980]{ric80} Richstone, D. O., \& 
	Schmidt, M. 1980, \apj, 235, 361 
\bibitem[Sanders et al. 1989]{san89} Sanders, D. B., Phinney, E. S., 
	Neugebauer, G., Soifer, B. T., \& Matthews, K. 1989, \apj, 347, 29
\bibitem[Sargent et al. 1989]{sar89} Sargent, W. L. W., Steidel, C. C., \& 
	Boksenberg, A. 1989, \apjs, 69, 703 
\bibitem[Serjeant \& Rawlings 1996]{ser96} Serjeant, S., \& Rawlings, S. 
	1996, Nature, 379, 304
\bibitem[Stone 1977]{sto77} Stone, R. P. S. 1977, \apj, 218, 767 
\bibitem[Webster et al. 1995]{web95} Webster, R. L., Francis, P. J., 
	Peterson, B. A., Drinkwater, M. J.,  
	\& Masci, F. J. 1995, Nature, 375, 469
\end{thebibliography}
\end{document}